\documentclass[10pt,twocolumn,letterpaper]{article}

\usepackage[pagenumbers]{iccv} 

\definecolor{iccvblue}{rgb}{0.21,0.49,0.74}
\usepackage[pagebackref,breaklinks,colorlinks,allcolors=iccvblue]{hyperref}

\def\paperID{*****}
\def\confName{ICCV}
\def\confYear{2025}

\title{Improving Multislice Electron Ptychography with a Generative Prior}

\author{Christian K. Belardi\footnotemark[1]~,
Chia-Hao Lee\footnotemark[1]~,
Yingheng Wang,
Justin Lovelace,\\
Kilian Q. Weinberger,
David A. Muller,
Carla P. Gomes\\
Cornell University\\
{\tt\small \{ckb73, cl2696, david.a.muller\}@cornell.edu, gomes@cs.cornell.edu}
}

\newcommand{\themethod}{\textsc{MEP-Diffusion}\xspace}

\begin{document}
\maketitle
\renewcommand{\thefootnote}{\fnsymbol{footnote}}
\footnotetext[1]{Equal Contribution}

\begin{abstract}
Multislice electron ptychography (MEP) is an inverse imaging technique that computationally reconstructs the highest-resolution images of atomic crystal structures from diffraction patterns. Available algorithms often solve this inverse problem iteratively but are both time consuming and produce suboptimal solutions due to their ill-posed nature. We develop \themethod, a diffusion model trained on a large database of crystal structures specifically for MEP to augment existing iterative solvers. \themethod is easily  integrated as a generative prior into existing reconstruction methods via Diffusion Posterior Sampling (DPS). We find that this hybrid approach greatly enhances the quality of the reconstructed 3D volumes, achieving a 90.50\% improvement in SSIM over existing methods.
\end{abstract}
\section{Introduction}
\label{sec:introduction}
Understanding the crystal structure of materials is pivotal for unraveling the fundamental interactions that govern the physical world. Electron microscopy has long been the cornerstone of atomic-scale imaging that provides unmatched spatial resolution. Among its advanced methodologies, \textit{multislice electron ptychography} (MEP) has emerged as a powerful computational imaging technique, capable of achieving sub-angstrom resolutions by formulating the multiple electron scattering process as an inverse problem~\cite{chen2021electron, zhu2025insights, kp2025electron}. This capability allows researchers to reconstruct atomic structures from diffraction patterns with unprecedented clarity.

However, the current state-of-the-art in electron ptychography faces critical challenges. While effective at resolving structures in the lateral (i.e., height and width) dimensions, the depth resolution is often limited to around 2 nm, or 100 times worse than the lateral resolutions due to experimental constraints~\cite{chen2021electron, terzoudis2023resolution}. Additionally, conventional iterative solvers are very time-consuming, and can take hours to days to return a solution, while the data acquisition only takes a few second~\cite{tate2016high, philipp2022very, zambon2023kite, ercius20244d}. Lastly, these solvers can struggle with noisy measurements and produce suboptimal solutions due to the inherent ill-posed nature of the inverse problem.

\begin{figure}[!t]
\begin{center}
\centerline{\includegraphics[width=\columnwidth]{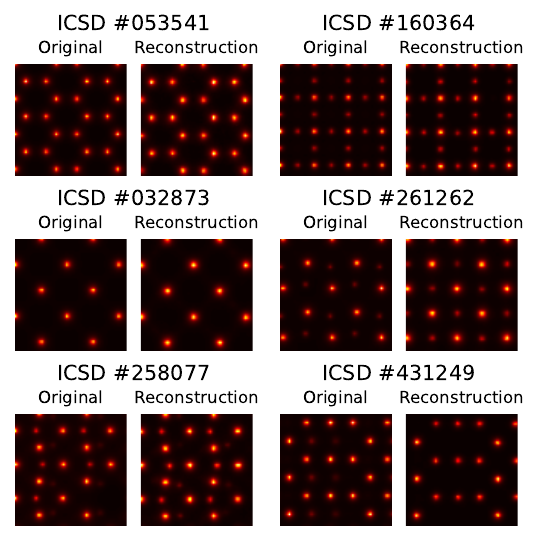}}
\caption{Side by side comparison of ground truth and \themethod reconstruction for selected crystal structure.}
\label{fig:teaser}
\end{center}
\vspace{-1cm}
\end{figure}

To address these limitations, we develop a generative prior using diffusion probabilistic models~\citep{ho2020denoising, song2020denoising, sohldickstein2015deep} and demonstrate its utility when combined with existing iterative solver. Diffusion models have demonstrated remarkable capability in various inverse problems by leveraging learned priors to produce high-quality reconstructions~\cite{chung2022diffusion, Feng_2023_ICCV}, and have been explored in Fourier ptychography~\cite{shamshad2019deep, wu2024fourier} and X-ray ptychography~\cite{cam2025ptychographic} for bulk specimens.
In this work, we introduce \themethod, which seamlessly integrates diffusion models into multislice electron ptychography. Our approach enables hybrid inverse problem solving, augmenting existing iterative solvers with a data driven model that captures the manifold of possible crystal structures. \autoref{fig:teaser} shows that \themethod produces high quality samples that match closely with the ground truth while remaining consistent with the physical measurements.

\begin{figure*}[!t]
\vskip 0.2in
\begin{center}
\centerline{\includegraphics[width=0.9\linewidth]{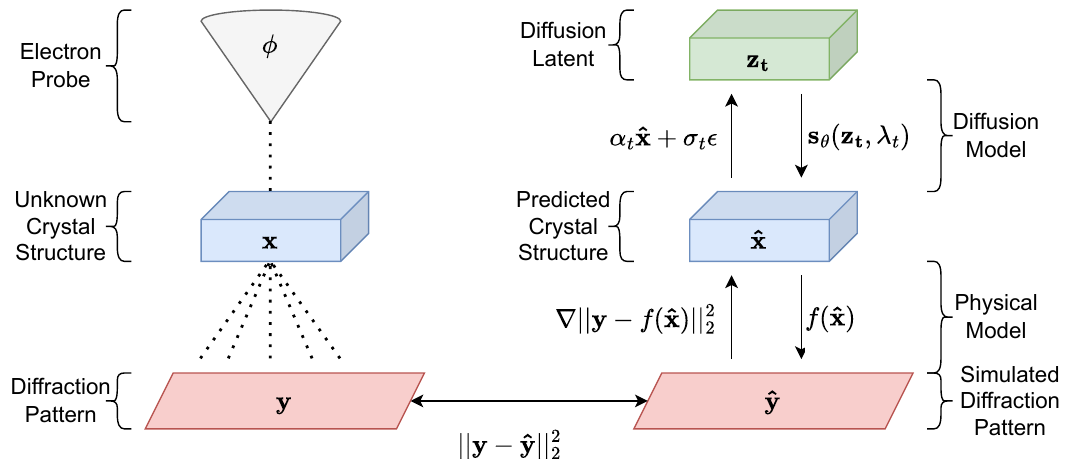}}
\caption{Overview connecting electron ptychography and diffusion posterior sampling. (Left) In the experimental setup, a focused electron probe is scanned across an unknown crystal structure on a 2D grid, recording a 2D diffraction pattern at each scan position. This results in a 4D dataset comprising a stack of 2D diffraction patterns over a 2D scan grid. (Right) \themethod reconstructs the crystal structure by combining a diffusion model prior with a differentiable physical model. The physical model guides the diffusion sampling process to ensure consistency between the generated structure and the observed diffraction patterns.}
\label{fig:setup}
\end{center}
\vskip -0.2in
\end{figure*}

\section{Preliminaries}
\label{sec:preliminaries}
\paragraph{Physical Forward Model}
In electron ptychography, we consider the measurement model in the following form:
$$\mathbf{y} = f(\mathbf{x}; \boldsymbol{\phi}, \boldsymbol{\Omega})$$
where $f$ represents the forward model, describing how the complex-valued electron probe $\boldsymbol{\phi}$ interacts with the material $\mathbf{x}$ at different probe positions $\boldsymbol{\Omega}$. The interaction produces a scattered wave function that propagates to a far-field detector, yielding real-valued diffraction patterns $\mathbf{y}$.

The materials are generally described by a complex-valued function $\mathbf{x}:=\mathbf{O}(\mathbf{r}) \in \mathbb{C}^{D \times H \times W}$, where \(\mathbf{r} = (x, y, z)\) represents spatial coordinates. Under the \textit{strong phase approximation (SPA)}~\cite{kirkland2010advanced}, $
    \mathbf{O}(\mathbf{r}) \approx \exp\left( i \sigma_e V(\mathbf{r}) \right)
$
where $i$ is the imaginary unit, $\sigma_e$ is the electron interaction parameter, $V(\mathbf{r})$ is the atomic scattering potential that captures the atom types and positions.

To simplify, we absorb the exponential function into the forward model, and redefine the crystal structure as $\mathbf{x} := \sigma_e V(\mathbf{r}) \in \mathbb{R}^{D \times H \times W}$. Because $\sigma_e$ is just a scaling constant that does not depend on $\mathbf{r}$, $\mathbf{x}$ contains all the structural information that electron ptychography aims to recover. By fixing both the probe $\boldsymbol{\phi}$ and the probe positions $\boldsymbol{\Omega}$, we arrive at a simplified forward model:
$$
    \hat{\mathbf{y}} = f(\mathbf{\hat{\mathbf{x}}}), \quad f: \mathbb{R}^{D \times H \times W} \rightarrow \mathbb{R}^{N_y \times N_x \times K_y \times K_x}
$$
where $N_y$ and $N_x$ correspond to different probe positions, while $K_y$ and $K_x$ denote the number of detector pixels.

The goal of electron ptychography is to retrieve the crystal structure $\mathbf{x}$ from diffraction patterns $\mathbf{y}$, which is commonly achieved by solving the following optimization problem:
$$
    \min_{\mathbf{x}} \quad \|\hat{\mathbf{y}} - \mathbf{y}\|_2^2 + \gamma \|\mathbf{x}\|_1
$$
where the first term ensures fidelity to the measured data, while the second term, which imposes sparsity in the solution and is controlled by the scalar weighting parameter $\gamma$ is optional, but commonly added to promote discrete atomicity within crystal structures.

We refer to this formulation as the \textit{electron ptychography reconstruction objective}. In prior work, this objective is typically minimized using first-order optimization methods, often requiring additional regularizations for stability.

\paragraph{Diffusion Probabilistic Models}
Given a dataset drawn from an unknown distribution $q(\mathbf{x})$, diffusion probabilistic models (DPMs)~\citep{sohl2015deep,ho2020denoising,song2020score, kingma2024understanding} learn a generative model $p_{\theta}(\mathbf{x})$ that approximates the data distribution $q(\mathbf{x})$. DPMs consist of a forward process and a generative process. The forward process defines a gradual transition between samples from the dataset distribution to samples from a Gaussian distribution. This introduces a series of increasingly noisy latent variables $\mathbf{z}_t$ for timesteps $t\in[0,1]$. For every timestep, the conditional $q_{t}(\mathbf{z}_t|\mathbf{x})$ is distributed such that $$\mathbf{z}_t = \alpha_t \mathbf{x} + \sigma_t \boldsymbol{\epsilon}, \text{ where } \boldsymbol{\epsilon} \sim \mathcal{N}(\mathbf{0}, \mathbf{I}).$$
Here, $\alpha_t,\sigma_t \in [0,1]$ are smooth scalar functions of the timestep $t$ which describe the \textit{noise schedule}. The noise schedule is defined such that the process starts with the clean data, $\mathbf{z}_0 \approx \mathbf{x} $, and the final latent is approximately Gaussian, $q(\mathbf{z}_1) \approx \mathcal{N}(\mathbf{0}, \mathbf{I})$. The diffusion process can equivalently be expressed in terms of the \textit{log signal-to-noise ratio} (logSNR), $\lambda_t = \log (\alpha_t^2 / \sigma_t^2)$, which decreases monotonically with $t$~\citep{kingma2021variational}. 

The generative process inverts the forward process, defining a gradual transition from samples of Gaussian noise to samples from the data distribution. Given access to the \textit{score function} $\nabla_{\mathbf{z}_t} \log q_t(\mathbf{z}_t)$, the forward process can be inverted exactly \citep{song2020score}. The key idea of diffusion models is to introduce a neural \textit{score network} to approximate the score function ${\mathbf{s_\theta}(\mathbf{z}_t, \lambda_t) \approx \nabla_{\mathbf{z}_t} \log q_t(\mathbf{z}_t)}$. The score network is trained with a denoising score matching objective over noise scales and data samples $\mathbf{x} \sim q(\mathbf{x})$:
$$
\mathcal{L}(\mathbf{x}) =  \mathbb{E}_{\boldsymbol{\epsilon} , t } \left[ w(\lambda_t) \|\mathbf{s_\theta}(\mathbf{z}_t, \lambda_t) - \nabla_{\mathbf{z}_t} \log q_t(\mathbf{z}_t|\mathbf{x})\|_2^2 \right],
$$
where $\mathbf{z}_t = \alpha_t \mathbf{x} + \sigma_t \boldsymbol{\epsilon}$.

In practice, the neural network can equivalently be parameterized as a noise ($ \boldsymbol{\epsilon} $) prediction, data ($ \mathbf{x} $) prediction, or velocity ($\mathbf{v} := \alpha_{t} \boldsymbol{\epsilon} - \sigma_{t} \mathbf{x}$) prediction network due to the following relationship
$$
    \nabla_{\mathbf{z}_t}\log q_t(\mathbf{z}_t|\mathbf{x}) = -\frac{\boldsymbol{\epsilon}}{\sigma_{t}}
    =- \frac{\mathbf{z}_{t}-\alpha_{t}\mathbf{x}}{\sigma_{t}^{2}}
    =-\mathbf{z}_{t} - \frac{\alpha_{t}}{\sigma_{t}}\mathbf{v}.
$$
These different parameterizations can be interpreted as imposing different weighting functions, $w(\lambda_t)$, over the denoising score matching objective and have been found to improve training stability and performance~\citep{ho2020denoising, kingma2024understanding}. We follow the best practices established in the image diffusion literature~\citep{kingma2024understanding} and parameterize our score network as a velocity prediction model, $\mathbf{v_\theta}(\mathbf{z}_t, \lambda_t)$~\citep{salimans2022progressive}.

\paragraph{Solving Inverse Problems with Diffusion Models}

During generation, we may want to impose some constraint over the generated data. For instance, we may have a partial measurement $\mathbf{y}$, and we want to draw samples $\mathbf{x}$ consistent with our measurement $\mathbf{y}$. To draw such samples, it is sufficient to approximate the conditional score function $\nabla_{\mathbf{z}_t} \log p_t(\mathbf{z}_t|\mathbf{y})$. We can decompose the conditional score function into the unconditional score and a likelihood term by Bayes' Rule: 
\begin{align*}
    \nabla_{\mathbf{z}_t} \log p_t(\mathbf{z}_t|\mathbf{y}) =\nabla_{\mathbf{z}_t} \log p_t(\mathbf{z}_t) + \nabla_{\mathbf{z}_t} \log p_t(\mathbf{y}|\mathbf{z}_t).
\end{align*}
This expansion shows that an unconditional diffusion model, which estimates $\nabla_{\mathbf{z}_t} \log p_t(\mathbf{z}_t)$, can incorporate additional constraints during generation via the likelihood term $ p_t(\mathbf{y} | \mathbf{z}_t)$. Critically, this is added during sampling and does not require re-training the diffusion network.

\paragraph{Diffusion Posterior Sampling}
For a variety of applications, we may have measurements for the \textit{clean} data $p(\mathbf{y}|\mathbf{x})$, but obtaining the distribution over \textit{noised} data $ p_t(\mathbf{y} | \mathbf{z}_t)$ is infeasible. In such cases, we perform DPS~\citep{chung2022diffusion} which uses the minimum mean squared estimate (MMSE) of the \textit{clean} data provided by the diffusion model, $\mathbf{x_\theta}(\mathbf{z}_t, t)$, and the likelihood term over the \textit{clean} data, $ p(\mathbf{y} | \mathbf{x})$, to approximate the conditional score function:
\begin{align*}
    \nabla_{\mathbf{z}_t} &\log p_t(\mathbf{z}_t|\mathbf{y}) \\
    &=\nabla_{\mathbf{z}_t} \log p_t(\mathbf{z}_t) + \nabla_{\mathbf{z}_t} \log p_t(\mathbf{y}|\mathbf{z}_t) \\
    &\approx\nabla_{\mathbf{z}_t} \log p_t(\mathbf{z}_t) + \nabla_{\mathbf{z}_t} \log p_t(\mathbf{y}|\mathbf{\mathbf{x}_\theta}(\mathbf{z}_t, \lambda_t)).
\end{align*}

This enables the direct incorporation of clean data measurements into the sampling process. At each sampling step, regardless of the parametrization of the diffusion model, we can compute the MMSE in closed form.

\begin{figure*}[ht]
\vskip 0.2in
\begin{center}
\centerline{\includegraphics[width=\linewidth]{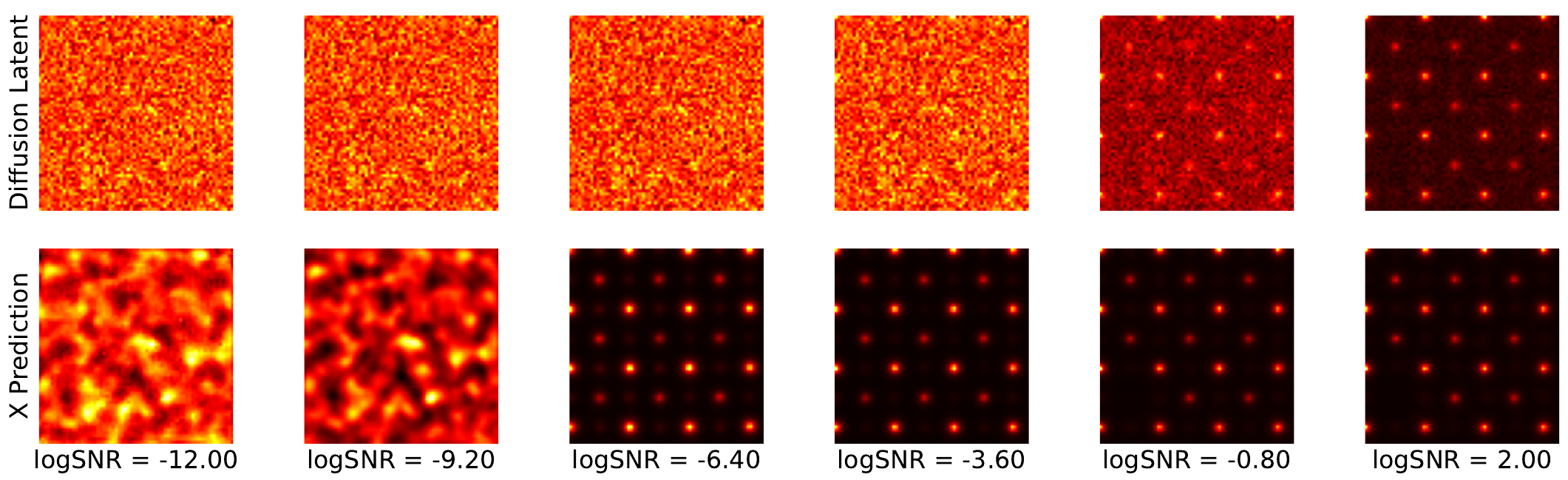}}
\caption{(Top) Visualization of the diffusion process across noise levels. (Bottom) Estimates of the original data by the trained diffusion network. We observe that periodic structure emerges very early in the diffusion process. Training at lower $\log\text{SNR}$ ranges compared to typical image diffusion models is critical for the model to learn a strong prior over atomic arrangements.}
\label{fig:decay-xpred}
\end{center}
\vskip -0.2in
\end{figure*}
\section{Ptychographic Reconstruction with DPMs}
\label{sec:methodology}

In electron ptychography, a focused electron probe is scanned across a material with an unknown crystal structure, producing a series of diffraction patterns. Our objective is to reconstruct the crystal structure that gave rise to these observed patterns. To accomplish this, we train a diffusion model to learn the underlying distribution of possible crystal structures and combine it with a differentiable physical model to sample crystal structures from the diffusion model that are physically consistent with the observed diffraction patterns. An overview of this framework is provided in~\autoref{fig:setup}.

\paragraph{DPMs for Crystal Structures}
We closely follow \citet{kingma2024understanding} in the design of our diffusion model, and parameterize the diffusion process in terms of the continuous $\log\text{SNR}$, denoted $\lambda_t$. We use a variance-preserving forward process, with $$\alpha_t^2 = \text{sigmoid}(\lambda_t), \quad \sigma_t^2 = \text{sigmoid}(-\lambda_t),$$ where $\lambda_t = \log(\alpha_{t}^2 / \sigma_{t}^2)$
is the $\log\text{SNR}$.

We optimize our network for the ``v''-prediction task proposed by \citet{salimans2022progressive} using a weighted mean squared error loss. 
Our training objective is therefore
$$
\mathbb{E}_{\boldsymbol{\epsilon} \sim \mathcal{N}(\mathbf{0}, \mathbf{I}), t \sim \mathcal{U}(0, 1)} \left[ w(\lambda_t) \, \|\mathbf{v_\theta}(\mathbf{z}_t, \lambda_t) - \mathbf{v}\|_2^2 \right].
$$

As shown in~\autoref{fig:decay-xpred}, we observe that periodic crystal structure emerges very early in the diffusion process at relatively low $\log\text{SNR}$. While diffusion models for natural images typically focus on moderate noise levels, we find it critical to concentrate training at higher noise levels where this periodic structure first emerges. See~\autoref{app:periodicity-and-logsnr} for more discussion.

Following \citet{karras2022elucidating}, we implement this using a Gaussian weighting in $\log\text{SNR}$ space: $w(\lambda_t) = \frac{1}{Z} \mathcal{N}(\lambda_t; -7, 3)$, where $Z$ is a normalization constant ensuring the density integrates to 1. This formulation centers the weighting at $\lambda_t = -7$, emphasizing high-noise regions during training. We find that shifting the loss weighting towards higher noise levels is critical for generating globally coherent samples. This observation aligns with recent findings in high-resolution image diffusion tasks~\citep{hoogeboom2023simple, chen2023importance, hoogeboom2024simpler}.
During training, we adopt the adaptive noise schedule from \citet{kingma2024understanding} to reduce variance.

Since the voxel intensities of the crystals are exponentially distributed, predicting $\mathbf{x}$ as the zero tensor is a strong baseline. To discourage this, we rescale our data so the standard deviation is approximately 1, corresponding to a constant scaling factor of $c=25$. This rescaling effectively up-weights the ``x''-prediction component of the ``v''-prediction task, for additional details see~\autoref{app:data-rescaling}.

Given the importance of capturing periodic structure at low $\log\text{SNR}$, we find that traditional image diffusion schedules like the cosine schedule perform poorly. Instead, we use a linear $\log\text{SNR}$ sampling schedule that concentrates steps in the range $[-13, 4]$, where periodicity emerges. This enables our model to dedicate more computation to the noise levels where atomic arrangements first appear. For improved sampling efficiency, we adopt the second-order DPMSolver++ sampler~\citep{lu2022dpm}, which outperforms first-order methods in our experiments.

\paragraph{DPS with the MEP Physical Model}
As shown in \citet{chung2022diffusion}, we can compute the score of the posterior from the sum of the scores of the prior and likelihood. The score of the prior is given by our unconditional diffusion model, \themethod. The score of the likelihood can be instantiated as a the loss of a classifier, or in our case the loss of the physical model.
$$\nabla_{\mathbf{z}_t}\log p_t(\mathbf{z}_t | \mathbf{y}) = \underbrace{\nabla_{\mathbf{z}_t} \log p_t(\mathbf{z}_t)}_{\text{Diffusion Model}} + \underbrace{\nabla_{\mathbf{z}_t} \log p_t(\mathbf{y} | \mathbf{z}_t)}_{\text{Physical Model Gradient}}.$$

\begin{figure}[h]
\begin{center}
\centerline{\includegraphics[width=\columnwidth]{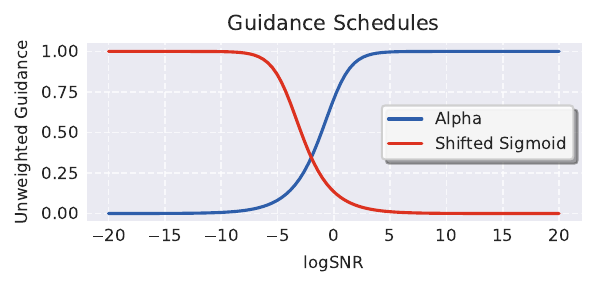}}
\caption{Illustration of guidance schedules: Alpha $g(\lambda_t) \propto \alpha_t$, and Shifted Sigmoid $g(\lambda_t) \propto \text{sigmoid}(4-\lambda_t)^{1/2}$}
\label{fig:guidance-schedule}
\end{center}
\end{figure}

Practically, following \citet{ho2022video}, we can implement this guidance as an adjustment to the ``x''-prediction:
    $$\mathbf{\mathbf{x}_{\theta}'}(\mathbf{z}_t, \lambda_t) = \mathbf{{\mathbf{x}}_{\theta}}(\mathbf{z}_t, \lambda_t) - g(\lambda_t) \nabla_{\mathbf{z}_t}||f(\mathbf{{\mathbf{x}}_{\theta}}(\mathbf{z}_t, \lambda_t)) - {\mathbf{y}}||_2^2$$
where $g(\lambda_t)$ is a guidance weight that controls the strength of the physical model contribution. While setting $g(\lambda_t) \propto \alpha_{t}$ yields ${{\mathbf{x}}_{\theta}'}(\mathbf{z}_{t}, \lambda_t) \approx \mathbb{E}_q[{\mathbf{x}}|{\mathbf{z}_{t}}, {\mathbf{y}}]$, we find that strong guidance towards the end of sampling leads to degenerate solutions with poor depth resolution (i.e., all the depth slices are nearly equivalent). We therefore anneal the guidance weight to smoothly reduce the influence of the physical model as sampling progresses, setting ${g(\lambda_t) \propto \text{sigmoid}(4-\lambda_t)^{1/2}}$. We visualize our proposed guidance schedule in~\autoref{fig:guidance-schedule}.

\begin{table*}[t]
\caption{Quantitative comparison of methods on instance-normalized reconstructions. We evaluate both full 3D reconstruction and depth-summed 2D projections using standard image quality metrics. Results show mean ± standard error of the mean across our test set. DPS gives statistically significant improvements ($p<1e-4$) over baseline methods on all metrics.}
\label{tab:quantitative}
\begin{center}
\begin{small}
\begin{sc}
\begin{tabular}{l|cc|cc}
\toprule
& \multicolumn{2}{c|}{Full Material Reconstruction} & \multicolumn{2}{c}{Material Depth Sum Reconstruction} \\
Method & PSNR $\uparrow$ & SSIM $\uparrow$ & PSNR $\uparrow$ & SSIM $\uparrow$ \\
\midrule
LSQ-ML & 27.4900 $\pm$ 0.1455 & 0.3949 $\pm$ 0.0090 & 27.9689 $\pm$ 0.1541 & 0.4279 $\pm$ 0.0096 \\
PtyRAD[Adam] & 26.8527 $\pm$ 0.1334 & 0.4629 $\pm$ 0.0104 & 27.4223 $\pm$ 0.1414 & 0.4896 $\pm$ 0.0108 \\
PtyRAD[L-BFGS] & 31.4885 $\pm$ 0.1501 & 0.6506 $\pm$ 0.0090 & 33.5574 $\pm$ 0.2003 & 0.7096 $\pm$ 0.0095 \\
\themethod & \textbf{33.7190 $\pm$ 0.1243} & \textbf{0.7523 $\pm$ 0.0053} & \textbf{41.5101 $\pm$ 0.1721} & \textbf{0.8852 $\pm$ 0.0041} \\
\bottomrule
\end{tabular}
\end{sc}
\end{small}
\end{center}
\end{table*}

Intuitively, the physical model guides the generation towards crystal structures consistent with the observed diffraction patterns while the diffusion model ensures atomic-scale details match the distribution of known crystal structures. This complementary approach is particularly effective for electron ptychography reconstruction - using physical guidance early in sampling establishes the correct periodic structure, while the learned prior later refines atomic-scale details that the physical model cannot resolve. This careful orchestration enables reconstruction of high-quality structures that are both physically consistent with measurements and structurally realistic at the atomic scale.

\paragraph{3D UNet with Anisotropic Resampling}
We modify the classic UNet architecture to accommodate the volumetric nature of crystal structures, as shown in~\autoref{fig:unet} of~\autoref{app:network-architecture}. The spatial sampling of our crystal data is anisotropic, with coarser sampling along the depth dimension (1.6~\AA) compared to the lateral dimensions (0.2~\AA). To handle this sampling anisotropy, we resample in two-stages. In the first stage, the UNet performs downsampling exclusively along height and width, reducing the feature spatial sampling from (1.6~\AA, 0.2~\AA, 0.2~\AA) to (1.6~\AA, 1.6~\AA, 1.6~\AA). In the second stage, downsampling is applied evenly across depth, height, and width. During upsampling, the process is reversed. First upsampling is applied uniformly across all dimensions and then only to the lateral dimensions, restoring the original anisotropic spatial sampling. Additional details are provided in~\autoref{app:network-architecture}.

\section{Experiments}
\label{sec:experiments}

\begin{figure*}[!t]
\vskip 0.2in
\begin{center}
\centerline{\includegraphics[width=0.9\linewidth]{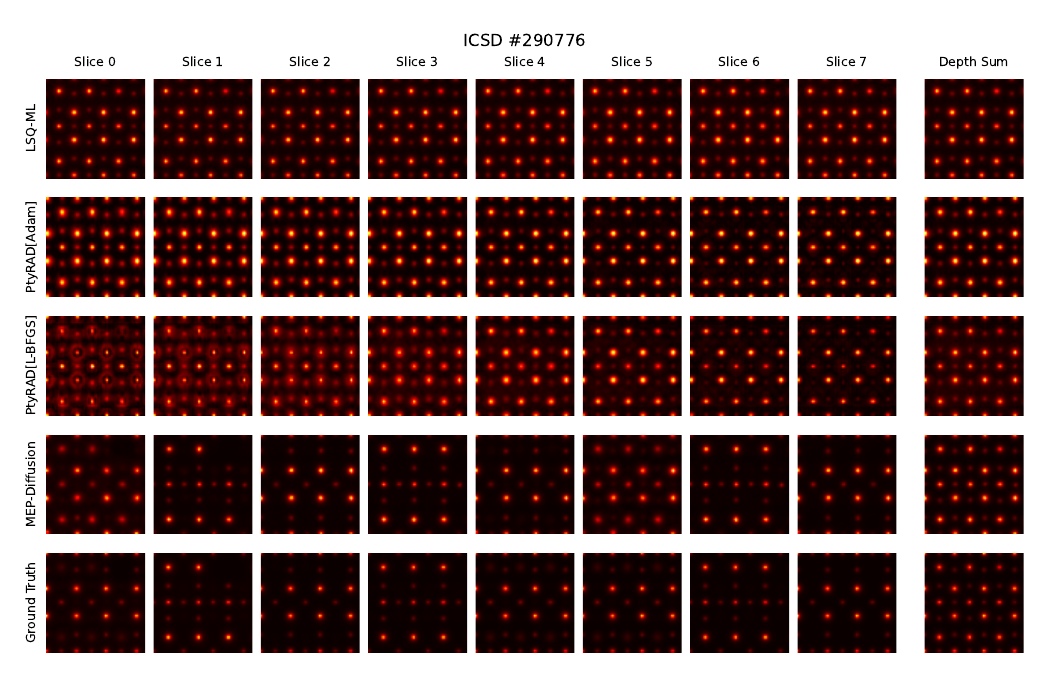}}
\caption{Qualitative comparison across depth slices of a crystal structure. Baseline methods struggle to recover depth variation, producing uniform reconstructions across slices. \themethod\ successfully reconstructs the full crystal structure.}
\label{fig:qualitative}
\end{center}
\vskip -0.2in
\end{figure*}

\begin{table*}[t]
\caption{Impact of our proposed Shifted Sigmoid guidance schedule. We evaluate both full 3D reconstruction (left) and depth-summed 2D projections (right) using standard image quality metrics. Results show mean ± standard error of the mean across our test set. Our guidance schedule leads to statistically significant improvements ($p<1e-4$) over the baseline schedule on all metrics.}
\label{tab:ablation}
\vskip 0.15in
\begin{center}
\begin{small}
\begin{sc}
\begin{tabular}{l|cc|cc}
\toprule
& \multicolumn{2}{c|}{Full Material Reconstruction} & \multicolumn{2}{c}{Material Depth Sum Reconstruction} \\
Guidance Schedule & PSNR $\uparrow$ & SSIM $\uparrow$ & PSNR $\uparrow$ & SSIM $\uparrow$ \\
\midrule
Alpha & 32.7677 $\pm$ 0.1126 & 0.6494 $\pm$ 0.0074 & 38.4307 $\pm$ 0.1517 & 
0.7745 $\pm$ 0.0061 \\
Shifted Sigmoid & \textbf{33.7190 $\pm$ 0.1243} & \textbf{0.7523 $\pm$ 0.0053} & \textbf{41.5101 $\pm$ 0.1721} & \textbf{0.8852 $\pm$ 0.0041} \\
\bottomrule
\end{tabular}
\end{sc}
\end{small}
\end{center}
\vskip -0.1in
\end{table*}

\paragraph{Dataset Preparation}
We generated train, validation, and test datasets using the crystallographic information files included in the Inorganic Crystal Structure Database (ICSD)~\citep{zagorac2019recent}. ICSD is the largest database for experimentally identified crystal structures and contains more than 258,000 entries that describe the atom types and relative positions, which provides informative prior knowledge of existing crystals. We filter the ICSD dataset so only the crystal structures that have all three lattice vectors ($|\vec{a}|$, $|\vec{b}|$, $|\vec{c}|$) less than 20~\AA\ are considered. After filtering we are left with $\approx 211,000$ unique materials each of which we render in three different orientations. We use abTEM~\citep{madsen2021abtem}, an electron microscopy simulation package, to render the crystal structures as 3-dimensional tensors. The final dataset contains $\approx$ 633,000 examples. 

We split off $5,000$ materials (i.e. $15,000$ examples) for our validation set, and another $100$ materials (i.e. $300$ examples) for our test set.
For the $15,000$ and $300$ examples in our validation and test sets, respectively, we simulate diffraction patterns using abTEM with realistic detector noise equivalent to an electron dose of $10^6$ $e^{-}$/{\AA}$^2$. These diffraction patterns will be used to evaluate the quality of different reconstruction methods. See~\autoref{app:dataset-details} for more details.

\paragraph{Evaluation}
We evaluate all methods on all $300$ pairs of $(\mathbf{x}, \mathbf{y})$ in our test set. That is, given a 4D diffraction dataset $\mathbf{y}$, recover the crystal structure $\mathbf{x}$. We quantitatively evaluate the reconstruction quality using two common distortion metrics: PSNR, and SSIM~\citep{wang2004image}.
We set the sampling steps to 2000 if no additional notes are given. Since different crystal structures naturally exhibit varying intensity scales due to their atomic composition and density, and phase retrieval techniques like ptychography are generally only accurate up to an arbitrary constant phase offset~\citep{rodenburg2019ptychography}, we apply instance normalization to each reconstruction so they range between [0,1] before computing metrics. This normalization focuses the evaluation on the relative spatial distribution of atomic densities rather than absolute intensity values, which can vary significantly between samples and weight differently without normalization. For our experiments, we perform paired bootstrap resampling \citep{koehn-2004-statistical} to quantify the statistical significance of our improvements. See~\autoref{app:hyperparameters} for training and sampling hyperparameters.

\paragraph{Baseline Methods}
We evaluate our approach against several baselines. Our primary comparison is with LSQ-ML \citep{odstrvcil2018iterative}, the current state-of-the-art algorithm for MEP. Additionally, we implement the same physical model using automatic differentiation to solve the \textit{electron ptychography reconstruction objective} with either Adam or L-BFGS. This baseline is implemented using an open-source reconstruction package, PtyRAD~\citep{lee2025ptyrad}, and is denoted as PtyRAD[Adam] or PtyRAD[L-BFGS] in our results~\citep{kingma2014adam, liu1989limited}. See~\autoref{app:ptyrad-physics-solver} for more details.

\paragraph{Reconstruction Performance}
We report reconstruction results across methods in~\autoref{tab:quantitative}. We visualize selected successful reconstructions in~\autoref{fig:teaser} and~\autoref{fig:qualitative}, as well as additional generations in~\autoref{app:additional-qualitative-comparisons}. We observe that \themethod\ quantitatively achieves more faithful reconstructions than all baselines across all metrics. The improvement over both baselines is statistically significant ($p<1e-4$) across all reconstruction metrics.

From our qualitative visualization, we observe that the baseline methods collapse to a uniform reconstruction across depth slices due to limitations of the physical model. \themethod\ benefits from its learned prior which helps to recover distinct reconstructions across the depth dimension that are occasionally better aligned with the ground truth material. This demonstrates the utility of a generative prior to push updates from the physical model towards realistic crystal structures.

\begin{figure}[t]
\vskip 0.2in
\begin{center}
\centerline{\includegraphics[width=\columnwidth]{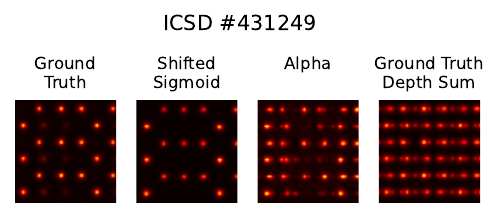}}
\caption{Qualitative comparison of our proposed Shifted Sigmoid guidance schedule and the baseline Alpha guidance schedule. Ground Truth, Shifted Sigmoid Guidance, and Alpha Guidance all display a single slice of the material, while Ground Truth Depth Sum shows the material summed over the depth dimension.}
\label{fig:guidance}
\end{center}
\vskip -0.2in
\end{figure}

\paragraph{Impact of Guidance Schedule} Our Shifted Sigmoid guidance schedule anneals the physical model influence to zero during generation, see~\autoref{fig:guidance-schedule}. This approach markedly differs from conventional guidance schedules in inverse problems, which typically increase guidance strength throughout generation. Without this carefully designed annealing, generation collapses along the depth dimension, exhibiting failure modes characteristic of direct optimization methods, see~\autoref{fig:guidance}. This collapse occurs because imprecise guidance from the physical model washes out the structure being formed by the diffusion model, forcing the generation toward degenerate solutions that satisfy physical constraints but are not realistic crystal structures. The Shifted Sigmoid schedule proves critical by gradually reducing physical guidance, enabling the diffusion model to resolve local atomic structure while still utilizing physical model guidance early in generation. The quantitative performance advantage persists across all metrics, see~\autoref{tab:ablation}. This finding highlights a fundamental difference from typical inverse problems studied in machine learning: the guidance models available for real-world applications are often imperfect or inadequate, unlike the near-perfect models commonly studied.

\begin{figure*}[!t]
\vskip 0.2in
\begin{center}
\centerline{\includegraphics[width=\linewidth]{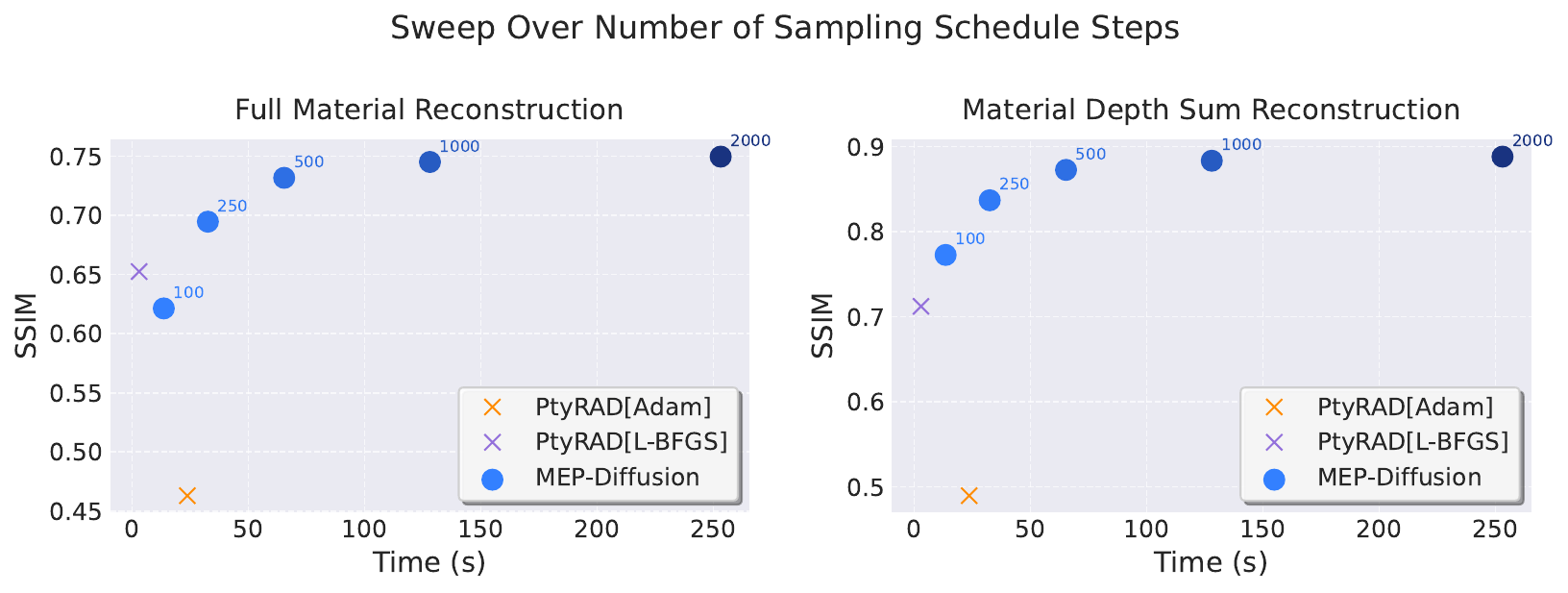}}
\caption{Trade-off between computational time and reconstruction quality. We evaluate \themethod\ across different numbers of sampling steps (100-2000) and compare against PtyRAD[Adam] and PtyRAD[L-BFGS] baselines. All time experiments were run on a single NVIDIA H100 GPU. Even with just 100 steps, \themethod\ achieves better reconstruction quality than baselines while maintaining reasonable computation time. Additional sampling steps provide further quality improvements at the cost of longer reconstruction times.}
\label{fig:performance-v-time}
\end{center}
\vskip -0.2in
\end{figure*}

\paragraph{Wall Clock Time Comparison}
We present the performance of \themethod\ across different sampling steps in~\autoref{fig:performance-v-time}. We observe that \themethod\ demonstrates strong performance even with relatively few sampling steps. While L-BFGS converges quickly to a suboptimal solution and Adam requires longer optimization, \themethod\ offers a flexible quality-to-time trade-off. With just 100 sampling steps, \themethod\ already outperforms both baselines in reconstruction quality. Increasing the number of steps yields further improvements, with diminishing returns beyond 1000 steps.

This flexibility is particularly valuable in practical applications, where different use cases may have different computational constraints. For rapid structure assessment, users can opt for fewer sampling steps while still obtaining better reconstructions than traditional methods. When reconstruction quality is paramount, additional sampling steps can be used to increase quality, albeit at increased computational cost. Importantly, this trade-off can be adjusted at inference time, without re-training the diffusion model.

\section{Related Work}
\label{sec:relatedwork}

\paragraph{Inverse Problems} 
Diffusion models have emerged as powerful tools for solving inverse problems through their learned priors, demonstrating effectiveness on natural image tasks like inpainting, super-resolution, and deblurring \citep{lugmayr2022repaint, chung2022mcg, chung2022diffusion, song2023pseudoinverseguided}, as well as diverse applications in medical imaging \citep{song2022solving}, motion synthesis \citep{zhang2023diffcollage}, and physical simulation \citep{fan2025neural}. While most prior work focuses on simple forward models, our work shows how to effectively incorporate complex physical models from electron microscopy while learning priors over periodic structures.

\paragraph{Incorporating Physical Constraints} Recent work has explored different approaches for incorporating physical constraints into deep generative models. Physics-informed neural networks (PINNs) incorporate general physical laws directly into the loss function during training \citep{raissi2017physics1, raissi2017physics2, raissi2019physics}. Past work has also explored training the diffusion model to respect physical constraints \citep{fan2025neural}. In contrast, our work tackles physical inverse problems by incorporating guidance from a differentiable physical forward model using measurements directly at \textit{inference} time.

\paragraph{Electron Ptychography} Traditional ptychographic reconstruction relies on direct methods \cite{bates1989sub, rodenburg1993experimental, odstrvcil2018iterative} or iterative algorithms \cite{maiden2009improved, odstrvcil2018iterative}, with the latter achieving higher resolution at the cost of long reconstruction times~\cite{chen2021electron, Nguyen2024achieving}. Recent deep learning approaches aim to improve efficiency by mapping diffraction measurements to reconstructions \cite{cherukara2020ai, hoidn2023physics, yao2022autophasenn, chang2023deep, friedrich2023phase, gan2024ptychodv}, offering near-instant feedback but with reduced quality and flexibility. Our work integrates a generative prior from crystal structures with iterative electron ptychography, enhancing reconstruction quality while preserving the adaptability of traditional methods.
\section{Limitations}
\label{sec:limitations}
\themethod significantly improves reconstruction quality over existing methods but has several shortcomings. Errors range from minor issues, such as slice misordering and missing or extra atoms, to more significant ones, including incorrect local structures and resolution loss. Most errors involve atomic displacement along the depth dimension, even though the depth sum image remains accurate. These issues likely arise from ambiguities in the physical forward model~\citep{terzoudis2023resolution}, making true atomic structure recovery uncertain, especially within the depth dimension. Additionally, the generative prior may contribute to some errors. We have included more qualitative examples in~\autoref{app:additional-qualitative-comparisons} to illustrate these errors.

\section{Future Directions}
\label{sec:futuredirections}
While our learning-based approach advances multislice electron ptychography (MEP), further improvements are needed. Specifically, enhancing robustness to weak guidance at low electron doses $10^3$ $e^{-}$/{\AA}$^2$ or lower is critical for dose-sensitive materials. Current methods struggle with depth reconstruction under these conditions. Additionally, developing principled methods for uncertainty quantification in atomic positions is crucial, particularly when noise is significant. Another important direction is improving the quantitativeness of reconstructions, enabling accurate recovery of atomic species and local chemical compositions from phase signals. Addressing these challenges could improve MEP and provide valuable insights for both machine learning applications and scientific discovery.
\section{Conclusion}
\label{sec:conclusion}
We present \themethod, a diffusion model trained on crystal structures, which we combine with an existing physical forward model through DPS for multislice electron ptychography reconstruction. Our approach addresses limitations of current ptychographic reconstruction methods: traditional optimization methods are slow and struggle with depth ambiguity.

We find the shifted sigmoid guidance schedule is essential for guided sampling in this specific domain. Sampling with the typical guidance schedules leads to degenerate solutions with poor depth resolution. By gradually reducing the physical model's influence during sampling, \themethod can recover realistic atomic-scale structures while maintaining consistency with measured diffraction patterns.

The combination achieves substantial quantitative improvements over existing methods, with statistically significant gains across all metrics. Most importantly, it successfully reconstructs distinct depth slices where baseline methods collapse to uniform solutions, addressing a key challenge in 3D ptychographic imaging. The method offers flexible quality-to-time trade-offs and performs well even with just 100 sampling steps. This work demonstrates that incorporating a generative prior over crystal structures into the reconstruction process can significantly improve ptychographic reconstruction quality.

\newpage
\section*{Acknowledgments}
CKB is supported by the National Science Foundation (NSF) through the NSF Research Traineeship (NRT) program under Grant No. 2345579. CHL is supported by the Eric and Wendy Schmidt AI in Science Postdoctoral Fellowship, a program of Schmidt Sciences, LLC. This work is also supported by the NSF through OAC-
2118310; the National Institute of Food and Agriculture (USDA/NIFA); the Air Force Office of Scientific Research (AFOSR); the New York Presbyterian Hospital; arXiv; and a Schmidt AI2050 Senior Fellowship.

{
    \small
    \bibliographystyle{ieeenat_fullname}
    \bibliography{main}
}
\newpage
\appendix
\onecolumn
\section{Dataset Details}
\label{app:dataset-details}
\paragraph{Dataset Generation}
 ICSD contains around 258,000 entries that are experimentally verified. We choose only the crystal structures with lattice constants (a, b, c) less than 20~\AA,because crystal structures with large lattice constants are relatively rare in nature. This produces a list of $\approx$ 211,000 materials. The CIF files for these selected materials are imported into abTEM, an electron microscopy simulation package, to render the crystal structure into 3-dimensional (3D) datacube sized (16, 100, 100) with the voxel sampling rate of (1.6~\AA, 0.2~\AA, 0.2~\AA) along depth, height, and width dimensions. This anisotropic voxel sampling is chosen to match with the electron ptychography resolution because typically the depth resolution is much poorer compared to lateral (height and width) resolution. We also randomly remove 1\% of the atoms to emulate vacancies that are commonly observed in real materials, this enhance the applicability of the generative model. The dataset is further augmented 3 times by orienting crystals along the 3 major a-, b-, and c-axis, resulting in a full dataset of $\approx$ 633,000 materials. The rendered 3D datacubes are converted from atomic electrostatic potential (volts) to phase change angle (radians) by taking the angle of the complex function \(\mathbf{O}(\mathbf{r}) \approx exp(i \sigma_e V(\mathbf{r}))\), so $\mathbf{x} := \sigma_e V(\mathbf{r}) = angle(\mathbf{O}(\mathbf{r})$), which is consistent with our electron ptychography forward model. The value range of such phase change angle is bounded between [-$\pi$, $\pi$] but commonly within [0, 1] because both $\sigma_e$ and $V(\mathbf{r})$ are larger than 0 and $\sigma_e V$ is usually less than 1 radian given our voxel sampling.

\paragraph{Training Set Augmentation}
During training, we first apply random in-plane rotation (depth dimension being the rotating axis), and then randomly crop a sub-cube sized (8, 64, 64). This produces a training datacube that spans across (12.8~\AA, 12.8~\AA, 12.8~\AA) in space and covers a couple repeating units cells.

\paragraph{Diffraction Pattern Simulation}
For \themethod\ sampling, we simulate diffraction patterns from test set materials using abTEM. The diffraction patterns are simulated with optical parameters comparable to actual experimental conditions, including 300 kV acceleration voltage for electrons, 21.4 semi-convergence angle, 200~{\AA} overfocus, 500 nm spherical aberration, 0.512~{\AA} scan step size, 26 $\times$ 26 scan pattern, and electron dose of $10^6$ $e^{-}$/{\AA}$^2$. We did not include partial coherence or phonon vibration because the effect will be quite limited at these experimental conditions.

\section{Data Rescaling}
\label{app:data-rescaling}
We demonstrate how data rescaling can be interpreted as up-weighting the ``x''-prediction component in the ``v''-prediction task. Consider the ground truth $\mathbf{v} = \alpha_t \boldsymbol{\epsilon} - \sigma_t \mathbf{x}$ and network prediction ${\mathbf{v}_{\theta}}(\mathbf{z}_t, t)$.
The original ``v''-prediction objective can be decomposed as:
\begin{align*}
\min_\theta ||\mathbf{v} - {\mathbf{v}_{\theta}}||_2^2 &= \min_\theta || (\alpha_t \boldsymbol{\epsilon} - \sigma_t \mathbf{x}) - (\alpha_t {\boldsymbol{\epsilon}_\theta} - \sigma_t {\mathbf{x}_\theta})||_2^2 \\
&= \min_\theta ||\alpha_t \boldsymbol{\epsilon} - \sigma_t \mathbf{x} - \alpha_t {\boldsymbol{\epsilon}_\theta} + \sigma_t {\mathbf{x}_\theta}||_2^2 \\
&= \min_\theta ||\alpha_t({\boldsymbol{\epsilon}} - {{\boldsymbol{\epsilon}}_\theta}) + \sigma_t({\mathbf{x}_\theta} - \mathbf{x})||_2^2
\end{align*}
When we rescale the data by a factor $c$ such that $\mathbf{x}' = c \mathbf{x}$, the new objective becomes:
\begin{align*}
\min_\theta ||\mathbf{v}' - {\mathbf{v}'_{\theta}}||_2^2 &= \min_\theta ||\alpha_t({\boldsymbol{\epsilon}} - {{\boldsymbol{\epsilon}}_\theta}) + c \sigma_t({\mathbf{x}_\theta} - \mathbf{x})||_2^2.
\end{align*}
This shows that the ``x''-prediction component is up-weighted by a factor of $c$ in the revised objective.

\section{PtyRAD Solver for Ptychographic Reconstruction}
\label{app:ptyrad-physics-solver}
We use PtyRAD~\cite{lee2025ptyrad}, an open-source ptychographic reconstruction package, to implement iterative gradient descent algorithms for computing gradients required by DPS and to reconstruct baselines with Adam L-BFGS optimizers. PtyRAD leverages PyTorch’s automatic differentiation engine to efficiently compute gradients of optimizable tensors. For all experiments, we first learn a single fixed probe using Adam, and hold it constant for all reconstruction methods using the physical forward model because the test data are simulated with the same probe condition. The probe is fit on 32 examples from our validation set. We minimize the \textit{electron ptychography reconstruction objective}, except we minimize both the 32 atomic structures and the probe simultaneously. For PtyRAD[Adam] and PtyRAD[L-BFGS] baselines, we adopt the mini-batch update scheme and use a batch size of 32 diffraction patterns for each update step. A learning rate of 5e-4 is used for Adam while L-BFGS is run three times with the following learning rates: 1, 1e-1, 1e-2. We select the best result from L-BFGS according to the loss as our baseline. The Adam baseline is reconstructed with 200 iterations, while the L-BFGS baseline is reconstructed with 5 iterations because it converges faster. Note that each iteration in the Adam baseline corresponds to a full pass of all 676 diffraction patterns per example, while in the L-BFGS baseline each iteration is done by evaluating 20 randomly chosen mini-batches to get the estimation of the Hessian matrix. The total number of diffraction patterns seen by each optimizer for each iteration is roughly the same.

\section{Network Architecture Details}
\label{app:network-architecture}
\begin{figure*}[ht]
\vskip 0.2in
\begin{center}
\centerline{\includegraphics[width=\linewidth]{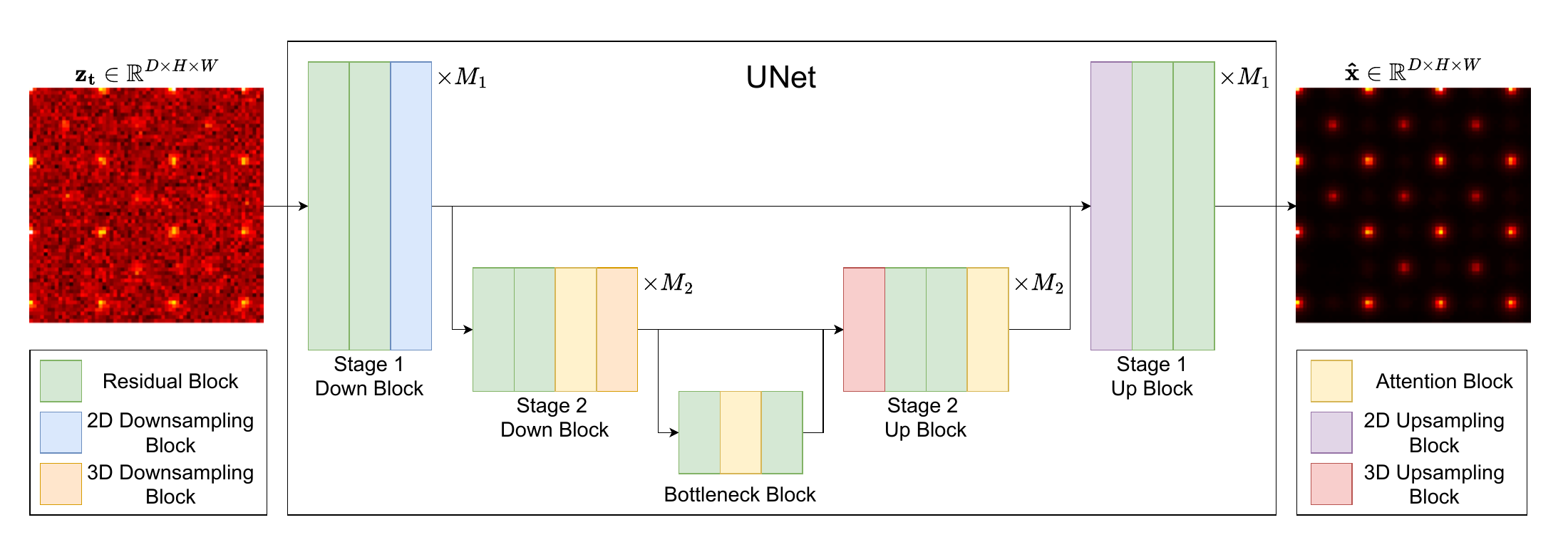}}
\caption{3D UNet architecture with anisotropic processing. The network first downsamples along height and width to match the coarser spatial sampling along depth, then performs downsampling across all dimensions. Self-attention layers are applied in low resolution blocks.}
\label{fig:unet}
\end{center}
\vskip -0.2in
\end{figure*}

The UNet architecture is composed of five macro structures, illustrated in~\autoref{fig:unet}.
\begin{itemize}
\item Stage 1 Down: Residual Block, Residual Block, 2D Downsampling/Convolutional Block
\item Stage 2 Down: Residual Block, Residual Block, Attention, 3D Downsampling/Convolutional Block
\item Bottleneck: Residual Block, Attention, Residual Block
\item Stage 2 Up: 3D Upsampling/Convolutional Block, Residual Block, Residual Block, Attention
\item Stage 1 Up: 2D Upsampling/Convolutional Block, Residual Block, Residual Block
\end{itemize}
The architecture only downsamples or upsamples the spatial dimensions when the channel count changes. Specifically, we downsample when increasing channel dimensions and upsample when decreasing them. Each down structure maintains two connections to its corresponding up structure.
The channel progression for each stage is as follows:
\begin{itemize}
\item Stage 1 Down: 16, 16, 32, 32, 64, 64
\item Stage 2 Down: 128, 128, 128, 128, 256, 256, 256, 256
\item Bottleneck: 256
\item Stage 2 Up: 256, 256, 256, 256, 128, 128, 128, 128
\item Stage 1 Up: 64, 64, 32, 32, 16, 16
\end{itemize}
The network's spatial transformation is significant: after Stage 1 Down, the input tensor of $8\times 64\times 64$ is compressed to $8\times 8\times 8$, with voxel sampling evolving from (1.6~\AA, 0.2~\AA, 0.2~\AA) to (1.6~\AA, 1.6~\AA, 1.6~\AA). Stage 2 Down performs uniform downsampling across all dimensions, with Stage 1 Up ultimately restoring the original tensor size and sampling rate. The model has 100M total parameters.

\section{Hyperparameters}
\label{app:hyperparameters}
See \autoref{tab:training-hyperparameters} for training hyperparameters, and \autoref{tab:sampling-hyperparameters} for sampling hyperparameters.

\begin{table}[!h]
\begin{minipage}{0.48\textwidth}
  \centering
  \begin{tabular}{@{}c|c@{}}
    \toprule
    Hyperparameter & Setting \\
    \midrule
    Optimizer & AdamW \\
    Batch Size & 256 \\
    Learning Rate & 0.0001 \\
    Weight Decay & 0.01 \\
    $\lambda_1$ & -20 \\
    $\lambda_0$ & 20 \\
    $w(\lambda_t)$ & $\frac{\mathcal{N}(\lambda_t;-7, 3)}{Z}$ \\
    \bottomrule
  \end{tabular}
  \caption{Table of Training Hyperparameters.}
  \label{tab:training-hyperparameters}
\end{minipage}%
\hfill
\begin{minipage}{0.48\textwidth}
  \centering
  \begin{tabular}{@{}c|c@{}}
    \toprule
    Hyperparameter & Setting \\
    \midrule
    Sampler & SDE DPMSolver++ \\
    Schedule & PolyExponential \\
    Min Sigma & 0.1 \\
    Max Sigma & 800 \\
    $\rho$ & 1.0 \\
    $g(\lambda_t)$ & $5000 *\text{sigmoid}(4 - \lambda_t)^{1/2}$ \\
    \bottomrule
  \end{tabular}
  \caption{Table of Sampling Hyperparameters.}
  \label{tab:sampling-hyperparameters}
\end{minipage}
\end{table}

\section{Periodicity and logSNR}
\label{app:periodicity-and-logsnr}
One of the critical adjustments required to fit a diffusion model on this periodic data was to tune our loss weighting to focus the model on very low logSNR regions. This is because as soon as the faintest crystal emerges the model immediately can recognize it. In order to unconditionally sample periodic crystals with our model we need the network to be capable of hallucinating periodicity from Gaussian noise and therefore we focus training on the region where the periodicity is beginning to emerge. We provide an example of the trained network predicting the crystal from diffusion latents at different logSNR in~\autoref{fig:decay-xpred}.

\section{Additional Qualitative Comparisons}
\label{app:additional-qualitative-comparisons}
\begin{figure*}[!t]
\vskip 0.2in
\begin{center}
\centerline{\includegraphics[width=0.8\columnwidth]{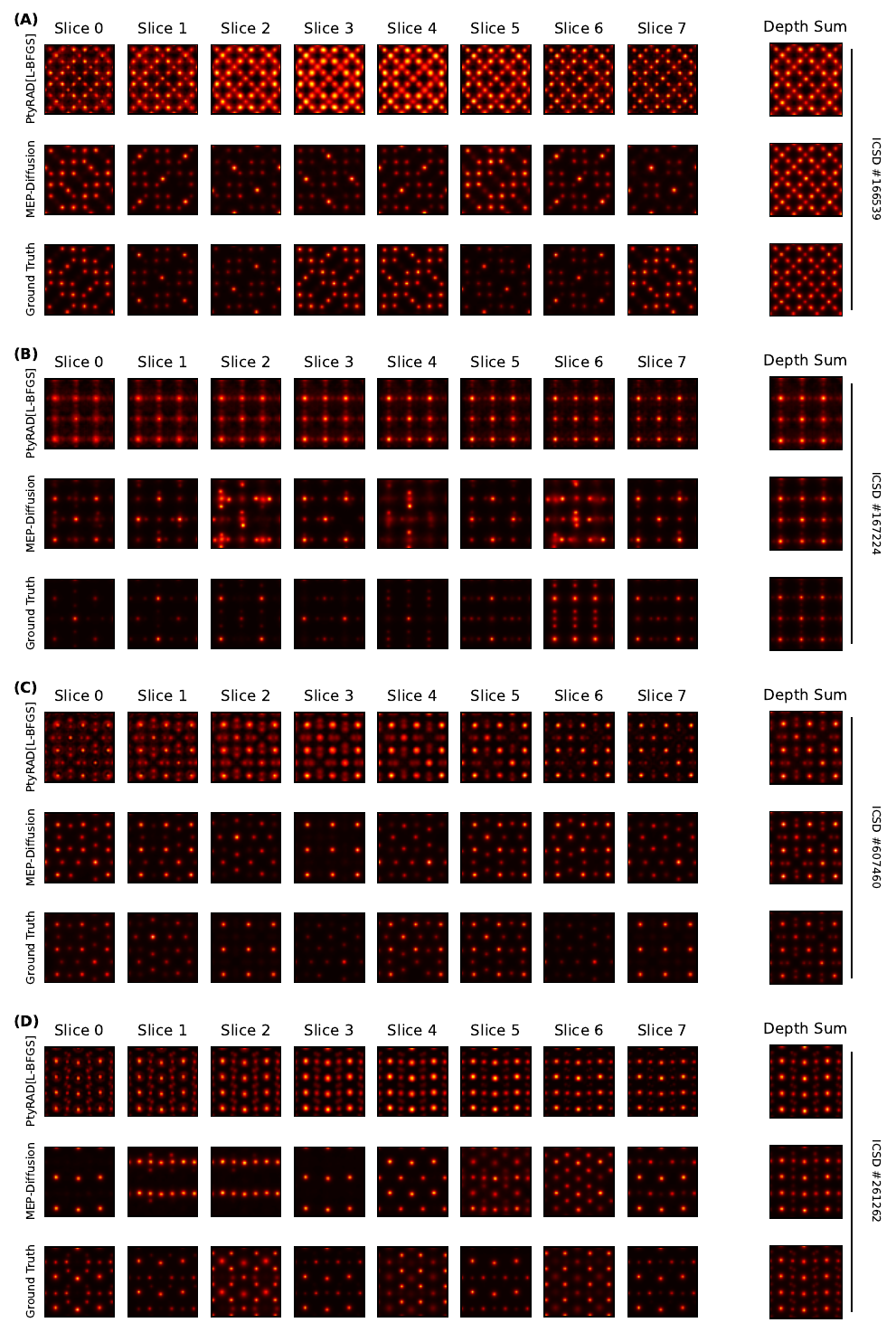}}
\caption{Additional qualitative comparisons.}
\label{fig:extra-qualitative}
\end{center}
\vskip -0.2in
\end{figure*}

\begin{figure*}[!t]
\vskip 0.2in
\begin{center}
\centerline{\includegraphics[width=0.8\columnwidth]{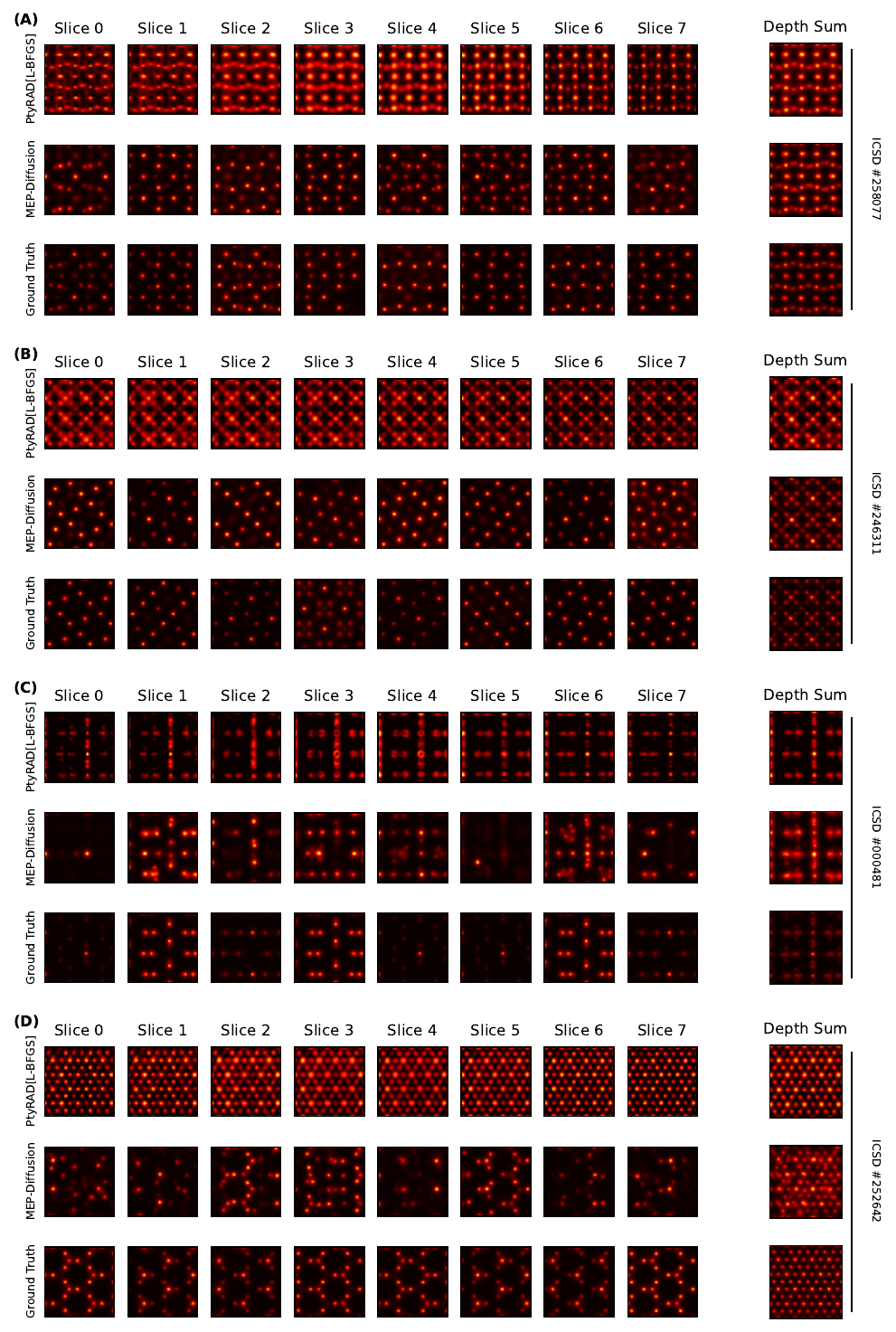}}
\caption{Additional qualitative comparisons.}
\label{fig:extra-qualitative2}
\end{center}
\vskip -0.2in
\end{figure*}

We include additional qualitative comparisons in~\autoref{fig:extra-qualitative} and~\autoref{fig:extra-qualitative2}.
We do this primarily to illustrate the limitations discussed in~\autoref{sec:limitations}. In~\autoref{fig:extra-qualitative}, subfigure (A) shows a near perfect reconstruction, subfigure (B) shows a generation where the structure is blurred, subfigure (C) shows a generation where many slices are correct but misordered, subfigure (D) shows a generation with incorrect structure. In~\autoref{fig:extra-qualitative2}, all subfigures show partially correct structures of varying degrees.

\end{document}